# Upper threshold for stability of multipole-mode solitons in nonlocal nonlinear media

Zhiyong Xu, Yaroslav V. Kartashov, and Lluis Torner

*ICFO-Institut de Ciencies Fotoniques and Universitat Politecnica de Catalunya,*

*08034, Barcelona, Spain*

We address the stability of multipole-mode solitons in nonlocal Kerr-type nonlinear media. Such solitons comprise several out-of-phase peaks packed together by the forces acting between them. We discover that dipole-, triple-, and quadrupole-mode solitons can be made stable, whereas all higher-order soliton bound states are unstable.

*OCIS codes: 190.5530, 190.4360, 060.1810*

The interactions arising between optical solitons generate a variety of phenomena. Unlike interactions of scalar solitons that tend to repel or attract each other depending on their relative phase difference only [1], the interaction between solitons incorporating several field components may be more complex. Thus, the formation of vector multipole-mode solitons is possible in local saturable [2,3] and in quadratic [4-6] media. The properties and interactions of solitons are also strongly affected by a nonlocality in the nonlinear response. Nonlocality is typical for photorefractive [1,2,7] and liquid [8,9] crystals; it is characteristic for thermal self-actions [10] and can be met in plasmas [11]. Nonlocality suppresses modulational instability of plane waves [12,13], and it can arrest collapse and instabilities of two-dimensional and vortex solitons (for a recent review see, e.g., Ref. [14]). Nonlocality also impacts soliton interactions [8,15] and allows formation of soliton bound states [6,15-17]. Dipole-mode bright solitons were observed in [18], while attraction of dark solitons was observed in [19]. However, the important issue of stability of bound states of bright solitons in nonlocal media was not addressed so far. In particular, the open question is: How many solitons can be packed into a stable bound state? In this Letter we report the outcome of such a stability analysis.

We consider the propagation of a slit laser beam along the $\xi$ axis in media with a nonlocal focusing Kerr-type nonlinearity described by the system of phenomenological equations for dimensionless complex light field amplitude $q$ and nonlinear correction to the refractive index $n$:

$$i\frac{\partial q}{\partial \xi} = -\frac{1}{2}\frac{\partial^2 q}{\partial \eta^2} - qn,$$

$$n - d\frac{\partial^2 n}{\partial \eta^2} = |q|^2, \quad (1)$$

where $\eta$ and $\xi$ stand for the transverse and the longitudinal coordinates scaled to the beam width and diffraction length, respectively; the parameter $d$ stands for the degree of nonlocality of the nonlinear response. When $d \to 0$ the Eq. (1) reduces to the single nonlinear Schrödinger equation; the case $d \to \infty$ corresponds to strongly nonlocal regime. Equations (1) describe nonlinear response of liquid crystals in steady state [8,9]. We neglect transient effects assuming continuous wave illumination (see [20] for a recent discussion of the reorientational relaxation time in typical crystals). Eq. (1) conserves the energy flow $U = \int_{-\infty}^{\infty} |q|^2 d\eta$ and Hamiltonian

$$H = \int_{-\infty}^{\infty} \left[ \frac{1}{2}\left|\frac{\partial q}{\partial \eta}\right|^2 - \frac{1}{2}|q|^2 \int_{-\infty}^{\infty} G(\eta-\lambda)|q(\lambda)|^2 d\lambda \right] d\eta, \text{ where } G(\eta) = (1/2d^{1/2})\exp(-|\eta|/d^{1/2})$$ is the response function of the nonlocal medium.

We search for stationary soliton solutions of Eq. (1) numerically in the form $q(\eta,\xi) = w(\eta)\exp(ib\xi)$, where $w(\eta)$ is the real function, $b$ is the propagation constant. To elucidate linear stability of soliton families we searched for perturbed solutions in the form $q(\eta,\xi) = [w(\eta) + u(\eta,\xi) + iv(\eta,\xi)]\exp(ib\xi)$, where real $u(\eta,\xi)$ and imaginary $v(\eta,\xi)$ parts of perturbation can grow with a complex rate $\delta$ upon propagation. Linearization of Eq. (1) around the stationary solution $w(\eta)$ yields the eigenvalue problem

$$\delta u = -\frac{1}{2}\frac{d^2 v}{d\eta^2} + bv - nv,$$
$$\delta v = \frac{1}{2}\frac{d^2 u}{d\eta^2} - bu + nu + w\Delta n,$$

(2)

where $\Delta n = 2\int_{-\infty}^{\infty} G(\eta-\lambda)w(\lambda)u(\lambda)d\lambda$ is the refractive index perturbation. We solved system (2) numerically.

First we recall properties of ground-state solitons (Fig. 1). The width of ground-state soliton increases while its peak amplitude decreases with increase of nonlocality degree $d$ at fixed $U$. The energy flow $U$ is a monotonically growing function of $b$ (Fig. 1(b)). As $b \to 0$ the soliton broadens drastically while its energy flow vanishes. Ground-state solitons are stable in the entire domain of their existence and realize the absolute minimum of Hamiltonian $H$ for a fixed energy flow $U$ (Fig. 1(c)).

The central result of this Letter is that several types of multipole-mode solitons can also be made completely stable in nonlocal media. Intuitively, multipole-mode solitons can be viewed as nonlinear combinations (bound states) of fundamental solitons with alternating phases. Such bound states cannot exist in local Kerr-type medium, where $\pi$ phase difference between solitons causes a local decrease of refractive index in the overlap region and results in repulsion. In contrast in nonlocal media, the refractive index change in the overlap region depends on the whole intensity distribution in transverse direction and under appropriate conditions the nonlocality can lead to an increase of refractive index and to attraction between solitons. The proper choice of separation between solitons results in bound state formation. Properties of simplest bound states of two solitons are summarized in Fig. 2. One can see that refractive index distribution features a small deep around the point $\eta = 0$ where light field vanishes (Fig. 2(a)). This deep is more pronounced at small nonlocality degree, while at $d \gg 1$ the refractive index distribution becomes almost bell-shaped. The energy flow of such solitons increases monotonically with increase of $b$ (Fig. 2(b)). At small energy flows dipole-mode solitons transform into two well-separated out-of-phase solitons, whose amplitudes decrease as $b$ decreases (Fig. 2(d)). The important result is that dipole-mode solitons are stable in the *entire* domain of their existence, even for small degrees of nonlocality $d \sim 0.1$ and at low energy levels, when solitons forming bound state are well separated (Fig. 2(d)).

Notice that bound soliton states were also studied in quadratic media, which can be regarded as nonlocal under appropriate conditions and can lead to similar equations for profiles of stationary solitons [6]. However, the principle difference between two systems becomes apparent upon stability analysis resulting in different eigenvalue problems and, hence, completely different stability properties of bound soliton states.

In order to answer the important question about the maximal number of solitons that can be incorporated into *stable* bound state we performed stability analysis of a number of higher-order soliton solutions. Results of stability analysis are summarized in Fig. 3. The energy flow of such solitons also grows monotonically with increase of $b$. In all cases in the regime of strong nonlocality $d \gg 1$ the refractive index distribution for multipole-mode soliton features bell-shaped profile with small modulation on its top in accordance with number of peaks in soliton (Figs. 3(a) and 3(c)). Stability analysis revealed that low-energy triple- and quadrupole-mode solitons are oscillatory unstable (Figs. 3(b) and 3(d)), but their complete stabilization is possible when soliton energy flow exceeds certain threshold. The width of instability domain as well as maximum growth rate decreases with increase of the nonlocality degree for both triple- and quadrupole-mode solitons (see, for example, Fig. 3(b)). It should be pointed out that at fixed $d$ the width of instability domain for triple-mode soliton is narrower than that for quadrupole-mode soliton (Fig. 3(d)). One of the most important results is that bound states incorporating five or more solitons were *all* found to be oscillatory unstable in the frames of the model (1) (see Fig. 3(e) and 3(f) for typical profile and dependence $\mathrm{Re}\,\delta(b)$ of unstable fifth-order soliton). We found this by performing linear stability analysis for bound states of up to 12 solitons and $d$ values from interval $(0,100)$. In all cases growth rate for unstable bound states was found to increase as $b \to \infty$ similarly to Fig. 3(f).

To confirm results of linear stability analysis, we performed numerical simulations of Eq. (1) with input conditions $q(\eta, \xi = 0) = w(\eta)[1 + \rho(\eta)]$, where $w(\eta)$ is the profile of stationary wave, $\rho(\eta)$ is a random function with a Gaussian distribution and variance $\sigma_{\mathrm{noise}}^2$. Stable dipole- and triple-mode solitons survive over huge distances in the presence of quite considerable broadband input noise (Fig. 4), while higher-order solitons self-destroy on propagation.

We also found that stability of bound soliton states is defined to a great extent by the character of nonlocal nonlinear response. Thus, in contrast to materials with exponential response function $G(\eta)=(1/2d^{1/2})\exp(-|\eta|/d^{1/2})$ produced by Eq (1) (as in liquid crystals), materials with Gaussian response function $G(\eta) = (\pi d)^{-1/2}\exp(-\eta^2/d)$ admit no upper threshold for number of solitons that can be incorporated into stable bound states. Such difference makes the search for materials with different characters of nonlocal response especially important.

Summarizing, we investigated the stability of multiple-mode solitons in focusing, nonlocal Kerr-type nonlinear media. We revealed that, in media with exponential nonlocal response, bound states are stable if they contain less than five solitons.

# References with titles

# References without titles

# Figure captions

Figure 1.  (a) Profile of ground-state soliton corresponding to point marked by circle in dispersion diagram (b) and Hamiltonian-energy diagram (c).

Figure 2.  (a) Profile of dipole-mode soliton corresponding to point marked by circle in dispersion diagram (b) and Hamiltonian-energy diagram (c). (d) Profile of low-energy dipole-mode soliton corresponding to $b = 0.13$ at $d = 5$.

Figure 3.  (a) Profile of triple-mode soliton at $b = 1.5$, $d = 5$. (b) Real part of perturbation growth rate for triple-mode soliton versus propagation constant. (c) Profile of quadrupole-mode soliton at $b = 2$, $d = 5$. (d) Real part of perturbation growth rate for triple- (1) and quarupole-mode (2) solitons versus propagation constant at $d = 5$. (e) Profile of fifth-order soliton at $b = 2$, $d = 8$. (f) Real part of perturbation growth rate for fifth-order soliton at $d = 8$.

Figure 4.  Stable propagation of dipole-mode (a) and triple-mode (b) solitons corresponding to $b = 1.5$ and $d = 5$ in the presence of white input noise with variance $\sigma^2_{\text{noise}} = 0.01$.

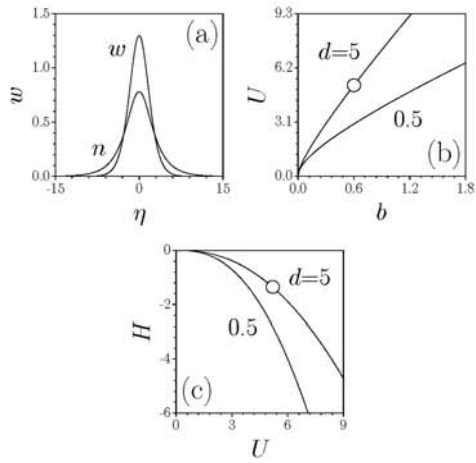

Figure 1. (a) Profile of ground-state soliton corresponding to point marked by circle in dispersion diagram (b) and Hamiltonian-energy diagram (c).

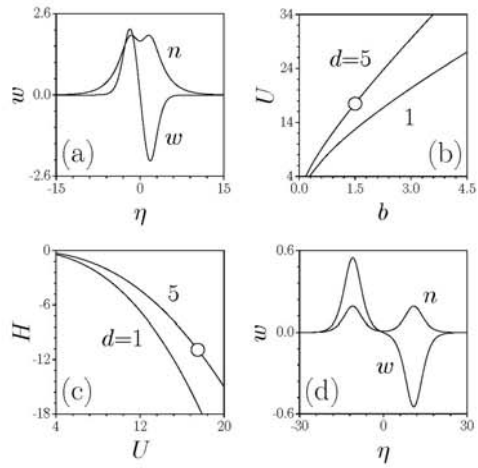

Figure 2.   (a) Profile of dipole-mode soliton corresponding to point marked by circle in dispersion diagram (b) and Hamiltonian-energy diagram (c). (d) Profile of low-energy dipole-mode soliton corresponding to $b = 0.13$ at $d = 5$.

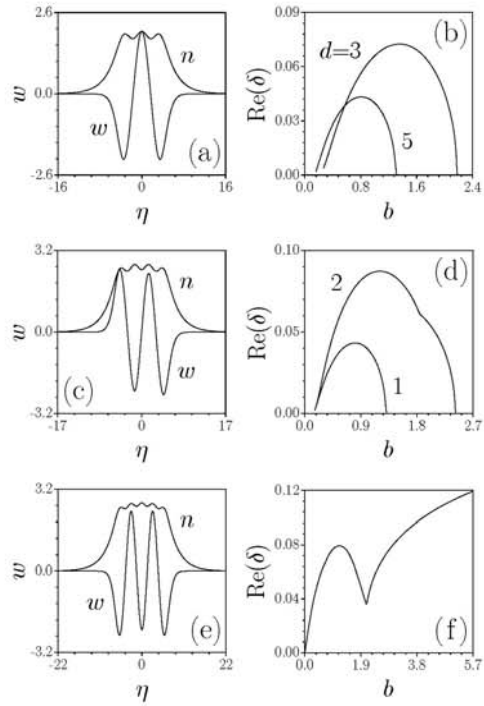

Figure 3. (a) Profile of triple-mode soliton at $b = 1.5$, $d = 5$. (b) Real part of perturbation growth rate for triple-mode soliton versus propagation constant. (c) Profile of quadrupole-mode soliton at $b = 2$, $d = 5$. (d) Real part of perturbation growth rate for triple- (1) and quarupole-mode (2) solitons versus propagation constant at $d = 5$. (e) Profile of fifth-order soliton at $b = 2$, $d = 8$. (f) Real part of perturbation growth rate for fifth-order soliton at $d = 8$.

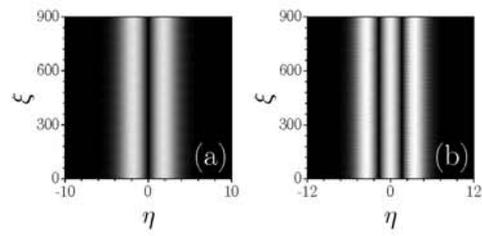

Figure 4. Stable propagation of dipole-mode (a) and triple-mode (b) solitons corresponding to $b = 1.5$ and $d = 5$ in the presence of white input noise with variance $\sigma^2_{\text{noise}} = 0.01$.